\documentclass[aps,pra,twocolumn,groupedaddress,showpacs]{revtex4}

\usepackage{graphicx}
\usepackage{amsmath}
\usepackage{bm}

\begin{document}


\title{Analogues of D-branes in Bose--Einstein Condensates}


\author{Kenichi Kasamatsu$^{1}$}
\email[]{kenichi@phys.kindai.ac.jp}
\author{Hiromitsu Takeuchi$^{2}$}
\author{Muneto Nitta$^{3}$}
\author{Makoto Tsubota$^{2}$}
\affiliation{
$^1$Department of Physics, Kinki University, Higashi-Osaka, 577-8502, Japan \\
$^2$Department of Physics, Osaka City University, Sumiyoshi-Ku, Osaka 558-8585, Japan \\
$^3$Department of Physics, and Research and Education Center for Natural 
Sciences, Keio University, Hiyoshi 4-1-1, Yokohama, Kanagawa 223-8521, Japan}


\date{\today}
\begin{abstract}
We demonstrate theoretically that analogues 
of D-branes in string theory can be realized in rotating, 
phase-separated, two-component Bose--Einstein condensates 
and that they are observable using current experimental techniques.
This study raises the possibility of simulating D-branes in the laboratory.
\end{abstract}

\maketitle

\section{Introduction}

String theory is the most promising candidate for producing a unified 
theory of the four fundamental forces of nature. 
Dirichlet (D-) branes, non-perturbative solitonic states of string theory, 
have been the most fundamental tool for studying non-perturbative dynamics 
in string theory. They are characterized as hypersurfaces 
on which open fundamental strings can terminate
with the Dirichlet boundary condition \cite{Polchinski:1995mt}.
D-branes are dynamic objects and their collective motion
is described by the Dirac--Born--Infeld (DBI) action in the low-energy regime \cite{Leigh}, 
which is a nonlinear action of the scalar field (corresponding to the transverse position
of the D-brane) and the $U(1)$ gauge field \cite{Dirac:1962iy,Born:1934gh}. 
Since the discovery of D-branes, string theory has developed in conjunction 
with the study of D-branes \cite{Polchinskibook}. 

Several years after the discovery of D-branes, a D-brane-like soliton was found 
in field theories such as the nonlinear sigma model 
(NL$\sigma$M) \cite{Gauntlett} and gauge theory \cite{Shifman},
in which vortex strings terminate on a domain wall between different vacua.
Here, the domain wall can be identified with a D-brane
in a sense that its collective motion is also described by the same DBI action as that in string theory, 
while vortex lines attached to it are analogous to 
fundamental strings because their endpoints are electrically charged 
and identical to solitons known as ``BIons" in the DBI action \cite{Gibbons:1997xz,Callan}. 
From this property, the authors of Ref. \cite{Gauntlett} called it ``D-brane soliton".
These theories thus offer simplified models for studying 
D-brane dynamics that are much easier to analyze than in full string theory.
All possible solutions of the wall-string composite
soliton have been classified and constructed in more general
sigma models and gauge theories \cite{Isozumi,Eto}. 

\begin{figure*}[t]
\begin{center}
\includegraphics[width=1.0\linewidth,keepaspectratio]{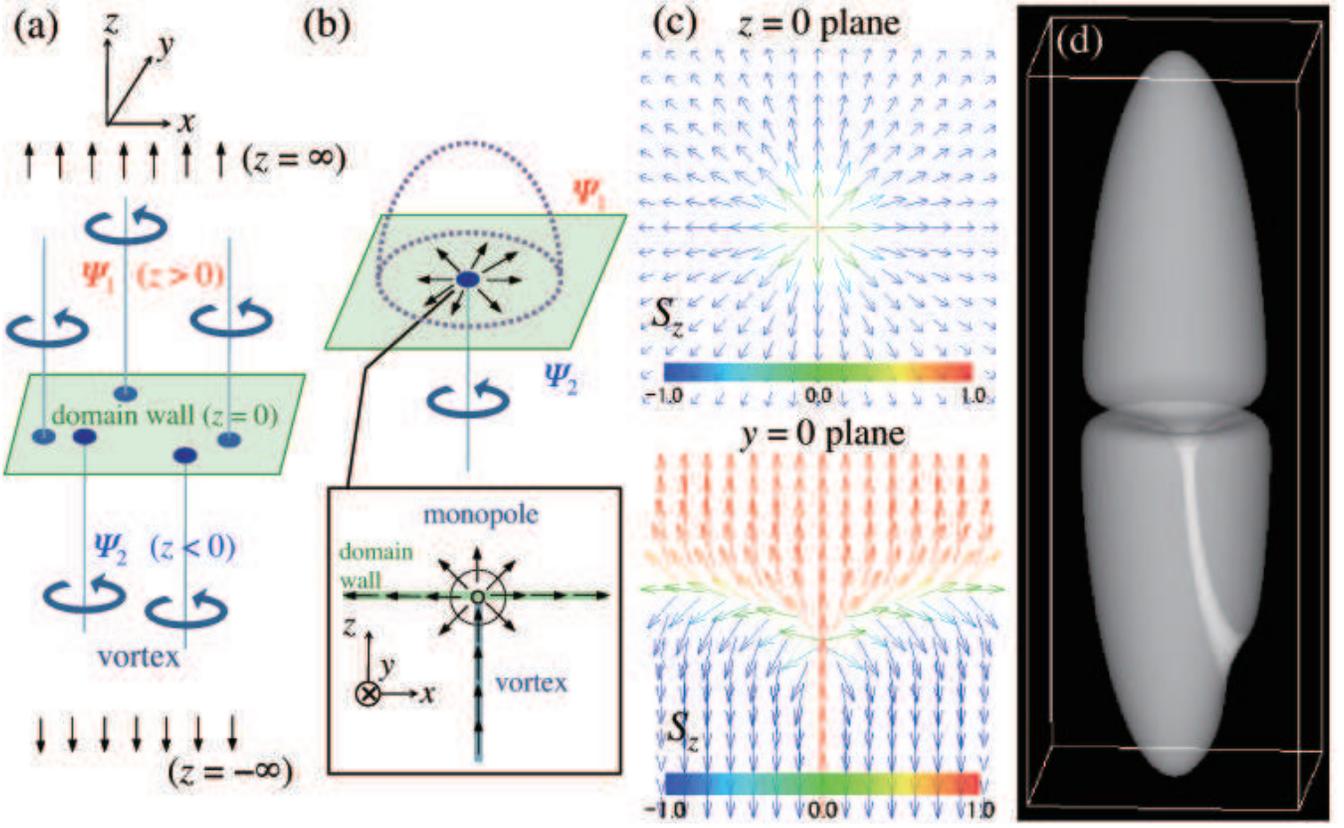}
\end{center}
\caption{D-brane soliton to which strings (vortices) attach in two-component BECs. 
(a) Schematic illustration of the wall--vortex soliton 
configuration viewed on a length scale larger 
than the domain-wall width and the vortex--core size. The two-component BECs 
$\Psi_{1}$ ($z>0$) and $\Psi_{2}$ ($z<0$) are separated by the domain 
wall in the $z=0$ plane. 
The boundary condition is given by $S_{z} \rightarrow 1$ ($-1$) for 
$z \rightarrow \infty$ ($-\infty$). 
We assume that vortex lines are straight and perpendicular to the wall. 
(b) The simplest D-brane soliton in which a single vortex along the $z$-axis for $z<0$ 
is connected to the domain wall. 
The spin texture (denoted by arrows) indicates that 
the connecting point can be identified as a monopole, 
as seen in the enlarged view of the region near the connecting point. 
(c) The spin texture ${\bf S}$ for the solution of Eq. (\ref{wallvortexcomp}) 
with $u_{\rm v}=1/\eta$, $\phi_{0}=0$, $z_{0}=0$, $M=1$, and the $+$ sign, 
in the $z=0$ and $y=0$ planes. 
The color indicates the magnitude of $S_{z}$. 
${\bf S} = \hat{\bf z}$ along the vortex core $(x=y=0)$.
(d) Equilibrium solutions for the phase-separated $^{87}$Rb--$^{85}$Rb BECs under rotation, 
obtained by numerical minimization of the energy functional for 
$\Omega = 0.38\omega_{x}$. 
The figure shows a normalized isosurface of the density difference 
$|n_{1}-n_{2}| b^{3}/N=1.5 \times 10^{-3}$ with the harmonic oscillator 
length $b=\sqrt{\hbar\omega_{x}/m}$ and the total particle number $N$. 
The vortex core appears as the lighter region. } 
\label{Dbrain1} 
\end{figure*}
Bose--Einstein condensates (BECs) of ultracold atoms are extremely 
flexible systems for studying solitons (or topological defects) since optical
techniques can be used to control and directly visualize the condensate 
wave functions \cite{Pethickbook}. Interest in various topological defects in BECs
with multicomponent order parameters has been increasing; the structure, stability, and creation/detection
schemes for monopoles \cite{Stoof,Martikainen,Savage}, three-dimensional 
Skyrmions \cite{Ruostekoski,Battye,Savage2}, 
and knots \cite{Kawaguchi} have been discussed.
In the present study, we consider a three-dimensional composite soliton in two-component BECs 
that consist of a domain wall and quantized vortices terminating on the wall, 
as sketched in Fig. 1(a). 
Specific examples of the system include a BEC mixture of two-species atoms 
such as $^{87}$Rb--$^{41}$K \cite{Thalhammer} or $^{85}$Rb--$^{87}$Rb \cite{Papp}, 
where the miscibility and immiscibility can be controlled by tuning 
the atom--atom interaction via Feshbach resonances. 
Here, the domain wall is referred to as an interface boundary of 
phase-separated two-component BECs. Although the interface has a finite thickness, the wall 
is well-defined as the plane in which both components have the same amplitude. 
Since a description of two-component BECs can be mapped to 
the NL$\sigma$M by introducing a pseudospin representation of the order 
parameter \cite{Kasamatsu2}, the resultant wall--vortex composite 
soliton corresponds to the D-brane soliton described in Ref. \cite{Gauntlett}. 
Here, the domain wall of the two components can be identified 
as a D-brane soliton because it has a localized $U(1)$ Nambu-Goldstone
mode which can be rewritten as $U(1)$ gauge field on the wall \cite{Gauntlett}, 
which is a necessary degree of freedom for the DBI action of a D-brane. 
In addition, vortex lines attached to the domain wall are identified as fundamental strings since 
their endpoints are electrically charged and identical to solitons called 
BIons in the DBI action. 
The primary differences from the D-brane soliton in Ref. \cite{Gauntlett} are only that our 
system is nonrelativistic and does not have supersymmetry. 
We find that these composite solitons are energetically stable 
in rotating, trapped BECs and are experimentally feasible with realistic parameters. 

\section{System}
The order parameter of two-component BECs is ${\bf \Psi} = (\Psi_{1}, \Psi_{2})$, 
where $\Psi_j=\sqrt{n_{j}}e^{i\theta_{j}}$ ($j=1,2$) are the macroscopically occupied spatial 
wave function of the two components with the density $n_{j}$ and  the phase $\theta_{j}$ . 
The order parameter can be represented by the pseudospin 
\begin{equation}
{\bf S} = (S_{x},S_{y},S_{z}) = (\sin \theta \cos \phi , \sin \theta \sin \phi , \cos \theta ) 
\end{equation}
with polar angle $\theta = \cos^{-1} [(n_{1}-n_{2})/n_{\rm T}]$ and azimuthal angle 
$\phi=\theta_{2}-\theta_{1}$ as 
\begin{equation}
{\bf \Psi}= \sqrt{n_{\rm T}} e^{i \frac{\Theta}{2}} \left( \cos \frac{\theta}{2} 
e^{-i \frac{\phi }{ 2}},  \sin \frac{\theta}{2} e^{i \frac{\phi }{ 2}} \right), 
\label{peudsospinrep}
\end{equation} 
where $n_{\rm T} = n_{1} + n_{2}$ and $\Theta = \theta_{1} + \theta_{2}$ represent the local density 
and phase, respectively \cite{Kasamatsu2}. The simplest wall--vortex configuration 
is schematically depicted in Fig. \ref{Dbrain1}(b), where 
the wall is characterized by the $S_{z} = 0$ plane and a vortex line in the $\Psi_{2}$ component 
along the $z$-axis attaches to it. Let us consider a surface in $z>0$ bound by the wall 
that encloses the end point of the vortex, shown by the dotted-curves. 
Then, the surface is mapped to a hemisphere in spin space, 
where the spin texture $(S_{x},S_{y}) = (\cos \phi , \sin \phi )$ on the wall winds 
once along the boundary.
The spin below the wall varies in a similar manner, 
except along the vortex line. Thus, the spin texture around the end point forms a 
monopole-like configuration. 

The solution of the solitonic structure in two-component BECs are given 
by the extreme of the Gross--Pitaevski (GP) energy functional 
\begin{eqnarray}
E [\Psi_{1},\Psi_{2}] = \int d^{3} r \biggl\{ \sum_{j = 1,2} 
\biggl[ \frac{\hbar^{2}}{2m_{j}}  \left| \left(\nabla - i \frac{m_{j}}{\hbar} {\bf \Omega} 
\times {\bf r} \right) \Psi_{j} \right|^{2} \nonumber \\ 
 + (\tilde{V}_{j} - \mu_{j}) |\Psi_{j}|^{2} 
+ \frac{g_{jj}}{2} |\Psi_{j}|^{4} \biggr] 
+ g_{12} |\Psi_{1}|^{2} |\Psi_{2}|^{2} \biggr\}.  \label{GPene}
\end{eqnarray}
Here, $m_{j}$ is the mass of the $j$th component and $\mu_j$ is its chemical potential. 
The system is supposed to rotate at the rotation frequency 
${\bf \Omega} = \Omega \hat{\bf z}$; thus, the harmonic trap potential 
$V_{j}=\frac{1}{2} m_{j} (\omega_{x}^{2} x^{2}+\omega_{y}^{2} y^{2}+\omega_{z}^{2} z^{2})$ 
is modified by the centrifugal term as
$\tilde{V}_{j} = V_{j} - m_{j}  \Omega^{2} (x^{2}+y^{2}) / 2 $. 
The coefficients $g_{11}$, $g_{22}$, and $g_{12}$ represent the 
atom--atom interactions. They are expressed in terms of the 
s-wave scattering lengths $a_{11}$ and $a_{22}$ 
between atoms in the same component and $a_{12}$ between atoms 
in different component as 
\begin{equation}
g_{jk} = \frac{2 \pi \hbar^{2} a_{jk}}{m_{jk}}
\end{equation}
with $m_{jk}^{-1} = m_{j}^{-1} + m_{k}^{-1}$. The GP model is given by the mean-field approximation 
for the many-body wave function and provides quantitatively good description of the 
static and dynamic properties of the dilute-gas BECs \cite{Pethickbook}.

\section{Mapping to the nonlinear sigma model}
To derive the generalized NL$\sigma$M for two-component BECs 
from the GP energy functional (\ref{GPene}), we assume $m_{1}=m_{2}=m$ and $V_{1}=V_{2}=V$. 
By substituting the pseudospin representation Eq.(\ref{peudsospinrep}) of ${\bf \Psi}$, 
we obtain \cite{Kasamatsu2} 
\begin{eqnarray}
E = \int  d {\bf r}  \biggl\{ \frac{\hbar^{2}}{2m} \biggl[ 
(\nabla \sqrt{n_{\rm T}})^{2}+ \frac{n_{\rm T}}{4} \sum_{\alpha} 
(\nabla S_{\alpha})^{2} \biggr] + V n_{\rm T} \nonumber \\
+ \frac{mn_{\rm T}}{2} ({\bf v}_{\rm eff} - {\bf \Omega} \times {\bf r})^{2}  
+  c_{0} + c_{1} S_{z} + c_{2} S_{z}^{2} \biggr\}, \label{gNLSM}
\end{eqnarray}
where we have introduced the effective superflow velocity
\begin{equation}
{\bf v}_{\rm eff} =\frac{ \hbar}{2m} ( \nabla \Theta - \cos \theta \nabla \phi ) 
\label{effvelociy}
\end{equation}
and the coefficients 
\begin{eqnarray}
c_{0} =\frac{ n_{\rm T}}{8} [n_{\rm T} (g_{11}+g_{22}+2g_{12}) - 4 (\mu_{1} + \mu_{2}) ], \\
c_{1} =\frac{ n_{\rm T}}{4}[ n_{\rm T} (g_{11}-g_{22}) - 2 (\mu_{1} - \mu_{2}) ], \\
c_{2} =\frac{ n_{\rm T}^{2}}{8} (g_{11}+g_{22}-2g_{12}). 
\end{eqnarray}
The coefficient $c_{1}$ can be interpreted as a longitudinal magnetic field 
that aligns the spin along the $z$-axis; it was assumed to be zero in this study. 
The term with the coefficient $c_{2}$ determines the spin--spin interaction 
associated with $S_{z}$; it is antiferromagnetic for $c_{2}>0$ and 
ferromagnetic for $c_{2}<0$ \cite{Kasamatsu2}.
Phase separation occurs for $c_{2}<0$, which we have focused on. 
Further simplification can be achieved by assuming that $V_{j}=0$ and the total density is uniform 
through the relation $n_{\rm T} = \mu / g$ where $g = g_{11} = g_{22}$ and 
$\mu = \mu_{1} = \mu_{2}$, and that the kinetic energy associated with the 
superflow ${\bf v}_{\rm eff} - {\bf \Omega} \times {\bf r}$ is negligible; 
the effects of these terms are discussed in the text. 
By using the healing length $\xi = \hbar/\sqrt{2mgn_{\rm T}}$ as the length scale, 
the total energy can reduce to 
\begin{eqnarray}
\tilde{E} = \frac{E}{g n_{\rm T} \xi^{3}} = \int  d {\bf r} \frac{1}{4} \left[ \sum_{\alpha} (\nabla S_{\alpha})^{2} 
+ M^{2} (1 - S_{z}^{2}) \right] \label{nonsigmamod2}, \\
M^{2} = \frac{4 |c_{2}|}{gn_{\rm T}^{2}},
\end{eqnarray}
where $M$ is the effective mass for $S_{z}$. This is a well-known massive 
NL$\sigma$M for effective description of a Heisenberg ferromagnet with spin--orbit coupling. 

The D-brane soliton by Gauntlett {\it et al}. \cite{Gauntlett} can be reproduced as follows. 
Introducing a stereographic coordinate $u=(S_{x}-iS_{y})/(1-S_{z})$, 
we can rewrite Eq. (\ref{nonsigmamod2}) as
\begin{equation}
\tilde{E} = \int d {\bf r} \frac{\sum_{\alpha} |\partial_{\alpha} u|^{2} +M^{2} |u|^{2}}
{(1+|u|^{2})^{2}}.
\label{nonsigmamod3}
\end{equation} 
For a fixed topological sector, vortices (a domain wall) parallel (perpendicular) to the $z$-axis, 
the total energy is bounded from below (Bogomol'nyi--Prasad--Sommerfield bound) as 
$\tilde{E} \geq |T_{\rm w}| + |T_{\rm v}|$ by the topological charges 
that characterize the wall and the vortex:
\begin{eqnarray}
T_{\rm w} = M \int  d {\bf r} \frac{ u^{\ast} \partial_{z} u + u \partial_{z} u^{\ast}}{(1+|u|^{2})^{2}}   ,  \\
T_{\rm v} = i \int  d {\bf r}  \frac{\partial_{x} u^{\ast} \partial_{y} u - \partial_{y} u^{\ast} \partial_{x} u }{(1+|u|^{2})^{2}} . 
\end{eqnarray}
Then, the equations 
\begin{equation}
\partial_{z} u \mp Mu =0, \hspace{6mm} (\partial_{x} \mp i \partial_{y}) u = 0 
\end{equation}
are satisfied, 
giving the analytic form of the wall--vortex composite solitons:
$u(z,\eta=x+iy)=u_{\rm w}(z) u_{\rm v}(\eta)$, where 
\begin{eqnarray}
u_{\rm w}(z)  = e^{\mp M (z-z_{0}) - i \phi_{0}}, \hspace{3mm}
u_{\rm v}(\eta) = 
\frac{\prod_{j=1}^{N_{v_1}} (\eta - \eta_{j}^{(1)})}{\prod_{j=1}^{N_{v_2}} (\eta - \eta_{j}^{(2)})}.
\label{wallvortexcomp}
\end{eqnarray}
The function $u_{\rm w}$ represents the domain wall with wall position $z_{0}$ 
and phase $\phi_{0}$ given by $(S_{x},S_{y})$; this phase $\phi_{0}$ yields the 
Nambu--Goldstone mode localized on the wall. 
The function $u_{\rm v}$ gives the vortex configuration, being written by arbitrary analytic functions 
of $\eta$; the numerator represents $N_{v_1}$ vortices 
in one domain ($\Psi_{1}$ component) and 
the denominator represents $N_{v_2}$ vortices in the 
other domain ($\Psi_{2}$ component). The positions of 
the vortices are denoted by $\eta_{j}^{(1)}$ and $\eta_{j}^{(2)}$.
In Ref. \cite{Gauntlett}, the solution is denoted by the coordinates $(X, \varphi)=(\tanh \log |u| , \arg u)$, 
where $X = \tanh [\mp M(z-z_{0}) + \log |u_{\rm v}(\eta) |]$. 
The total energy does not depend on the form of the solution, 
but only on the topological charges 
as $T_{\rm w} = \pm M$ or 0 (per unit area), and $T_{\rm v} = 2 \pi N_{\rm v}$ 
(per unit length), where 
$N_{\rm v}$ is the number of vortices passing through a certain $z=$ const plane.

\begin{figure*}
\begin{center}
\includegraphics[width=0.98 \linewidth,keepaspectratio]{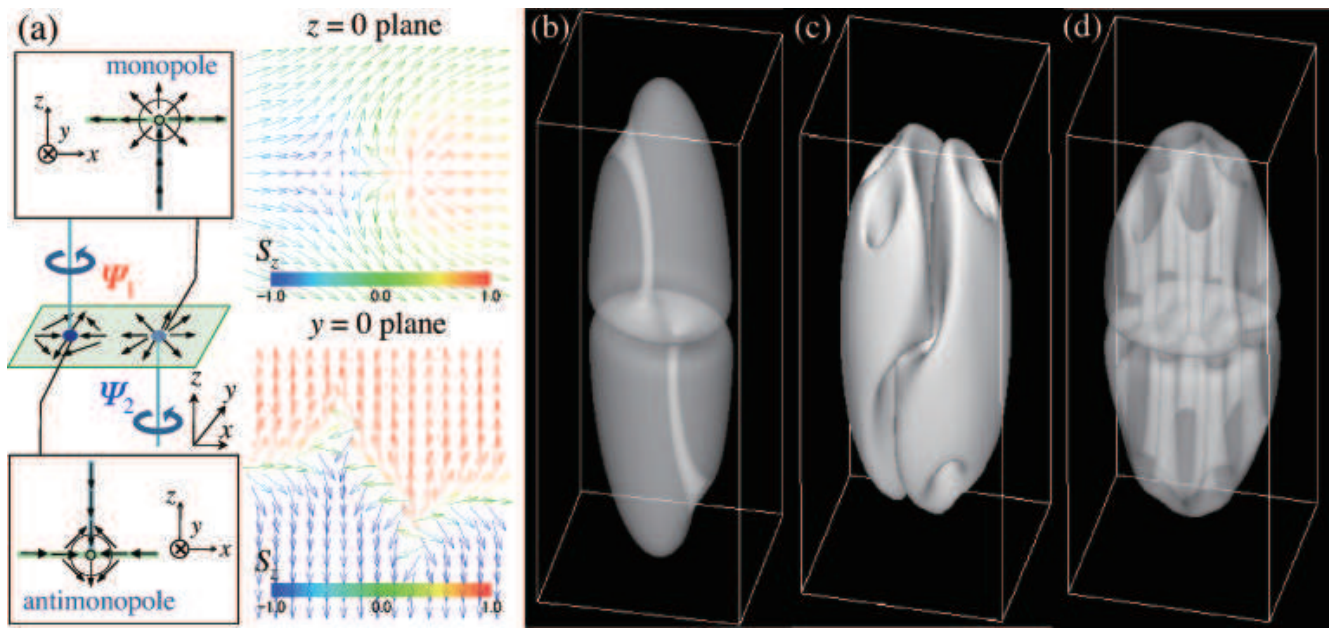}
\end{center}
\caption{D-brane to which several strings attach in two-component BECs. 
(a) Left: Schematic illustration of the configuration in which 
each component has a single vortex connected to the wall. 
The connecting points can be seen to be a monopole and an antimonopole. 
Right: The spin texture of this configuration in the $z=0$ and $y=0$ planes, 
using Eq. (\ref{wallvortexcomp}) with $u_{\rm v}=(\eta-x_{0})/(\eta+x_{0})$ and $x_{0} = \pm 2 $. 
The wall becomes asymptotically flat due to the balance 
between the tensions of the attached vortices. 
(b)--(d) Equilibrium solutions obtained by numerical minimization 
of the Gross--Pitaevskii energy functional for $^{87}$Rb--$^{85}$Rb BECs. 
The expression is the same as that in Fig. \ref{Dbrain1}(d).
The rotation frequency is (b) $\Omega = 0.40\omega_{x}$, and
(c) (d) $\Omega = 0.80\omega_{x}$. Only for (d), $a_{12} = 473 a_{0}$.} 
\label{fig:2} 
\end{figure*}

Figure \ref{Dbrain1}(c) shows the texture of ${\bf S}$ with the simplest wall--vortex 
configuration [Fig. \ref{Dbrain1}(b)], corresponding to the solution $u(z,\eta)$.
A vortex exists in $z<0$ and forms a texture known as the {\it lump} in field theory \cite{Belavin} 
or the Anderson--Toulouse vortex in superfluid $^3$He \cite{AndersonToulouse}, 
where the spin points up at the center and rotates continuously from up to down as it moves radially outward. 
The vortex ending attaches to the wall, causing it to bend 
logarithmically as $z=\log r/M$ [Fig. \ref{Dbrain1}(c) bottom]. 
We can construct solutions in which an arbitrary number of 
vortices are connected to the domain wall by multiplying by the additional factors
$\eta-\eta_{j}^{(i)}$ [see Eq. (\ref{wallvortexcomp})]; Fig. \ref{fig:2}(a) shows a solution in 
which both components have one vortex connected to the wall. 
In the NL$\sigma$M, the energy is independent of the vortex positions $\eta_{j}^{(i)}$ 
on the domain wall; in other words, there is no static interaction between vortices. 

Although analogues of ``branes" have already been 
studied in the AB phase boundary of superfluid $^{3}$He \cite{Bradley}, 
their exact correspondence to those in string theory has not been clarified. 
In our case, the domain wall 
has a localized $U(1)$ Nambu--Goldstone mode 
and it can be rewritten as the $U(1)$ gauge field on the wall, 
which is a necessary degree of freedom for the DBI action of a D-brane. 
Gauntlett {\it et al.} have shown that Eq. (\ref{wallvortexcomp}) reproduces 
the ``BIon" solutions of the DBI action for D-branes in string theory \cite{Gauntlett}, 
as can be demonstrated by constructing an effective theory of the domain wall world 
volume with collective coordinates $z_{0}(x,y)$ and $\phi_{0}(x,y)$ in $u_{\rm w}(z)$. 
On the domain wall, Eq. (\ref{wallvortexcomp}) 
becomes $M X = \sum_{j=1}^{N_{k_1}} \log (\eta - \eta_{j}^{(1)}) 
- \sum_{j=1}^{N_{k_2}} \log (\eta - \eta_{j}^{(2)})$ 
with $X=z_{0}+i\phi_{0}/M$ \cite{Sakai}. For the example shown in Fig. \ref{Dbrain1}(a), as we travel once 
around infinity $\eta \rightarrow \eta e^{2\pi i}$, the phase angle on the domain wall world volume 
shifts as $\phi_{0} \rightarrow \phi_{0}+2\pi$. When we introduce the $U(1)$ gauge field $A_{j}$ by 
taking a dual as 
\begin{equation}
\partial_{i} \phi_{0} = \epsilon_{ijk} \partial_{j} A_{k}, 
\end{equation}
the endpoints of the vortex strings are electrically charged particles \cite{Gibbons:1997xz,Callan}. 
Therefore, our domain wall can be identified as a D-brane on which fundamental strings terminate. 

\section{Numerical simulations}
To see whether the composite solitons in Eq. (\ref{wallvortexcomp}) are stable under actual experimental 
conditions, we study the wall--vortex soliton in a realistic setup of trapped two-component BECs 
by numerically minimizing the GP energy functional Eq. (\ref{GPene}) 
(equivalently, the generalized NL$\sigma$M Eq. (\ref{gNLSM}))
in the three-dimensional system via imaginary time propagation. 
Then, there are additional contributions to the massive NL$\sigma$M: a trapping potential, 
a gradient of $n_{\rm T}$, and a kinetic energy of superflow Eq. (\ref{effvelociy}) given by the gradients of $\Theta$ and ${\bf S}$.
According to Papp {\it et al.} \cite{Papp}, we set $\Psi_{1}$ ($\Psi_{2}$) to a $^{87}$Rb BEC ($^{85}$Rb BEC), 
and set the particle number to $N=5 \times 10^{4}$ for both components, 
the intraspecies s-wave scattering lengths to $a_{1}=a_{2}=100 a_{0}$, and 
the interspecies s-wave scattering length to $a_{12} = 213 a_{0}$, where $a_{0}$ is the Bohr radius and 
where the condition for phase separation $a_{12}^{2} > a_{1}a_{2}$ is satisfied \cite{Tim}. 
The experiment revealed that the value of $a_{2}$ was tuned in a wide range 
$50 a_{0}-900 a_{0}$ via Feshbach resonances \cite{Papp}. 
We prepare a cigar-shaped harmonic trap with frequencies $\omega_{x} = \omega_{y} = 20 \times 2 \pi$ Hz 
and $\omega_{z} = 5 \times 2 \pi$. Rotation ${\bf \Omega} = \Omega \hat{\bf z}$ is applied to stabilize 
vortices in the condensates.  

Figure \ref{Dbrain1}(d) shows the isosurface of the density difference $|n_{1}-n_{2}| \propto |S_{z}|$ 
of the stationary solution for $\Omega=0.38\omega$, 
representing the wall--vortex soliton corresponding to Fig. \ref{Dbrain1}(b) for trapped BECs; 
the regions of the domain wall ($S_{z} \simeq 0$) 
and the vortex core are clearly visible in this figure. 
The vortex in $\Psi_{2}$ forms a coreless vortex, where its core 
is filled by the density of $\Psi_{1}$ and transforms 
into a singular vortex with increasing distance from the domain wall. 
This configuration is energetically stable since it is obtained 
by imaginary time propagation. The spin texture of this solution is almost identical 
to that in Fig. \ref{Dbrain1}(c), despite there being extra contributions 
in the generalized NL$\sigma$M. Since the divergent kinetic energy generated by the vortex 
significantly depletes $n_{\rm T}$ to form a singular vortex core, it slightly modifies the potential of the spin field.

However, when the system contains multiple vortices, 
the above effects of the extra terms
become more important. Figure \ref{fig:2}(b) shows the equilibrium solution 
in which both components have a single vortex. The end point of the 
vortices in each component is spontaneously displaced from the center, 
while the energy is independent of $\eta_{j}^{(i)}$ for Eq. (\ref{wallvortexcomp}) 
given by the NL$\sigma$M. 
This is due to the effective repulsion between the endpoints of each vortex, 
originated from the two energetic constraints: a broad distribution of vorticity
near the domain wall to reduce the associated kinetic energy and 
a smooth distribution of $n_{\rm T}$ to reduce its gradient. 
When the rotation is further increased, multiple vortices are generated. 
For $\Omega=0.80\omega$ in our parameter setting, the domain wall tilts to be in 
parallel with the rotation axis, where some of the vortex lines are 
absorbed by the wall to form a ``vortex sheet" \cite{Kasamatsus} [Fig. \ref{fig:2}(c)]. 
To keep the domain wall perpendicular to the rotation axis for the fast 
rotation, one must increase the interspecies 
scattering length $a_{12}$, which decreases the interface area to reduce its energetic cost. 
Then, as shown in Fig. \ref{fig:2}(d), the vortex endings are also shifted relative to each other 
to form an interlaced rectangular lattice on the domain wall. 

\section{Conclusion and discussion}
We have shown that an analogue of a D-brane can be realized as an energetically stable solitonic 
object in phase-separated, rotating, two-component BECs. 
This suggests that atomic BECs are the most promising candidates for demonstrating D-brane physics 
in the laboratory; the energetically stable D-brane solitons warrant studying various dynamic phenomena, 
e.g., oscillation modes of strings and branes and nonlinear dynamics such as brane--antibrane annihilation, 
which was proposed as a possible explanation for the inflationary universe in string theory.
Here, we summarize the outlook gained by this realization.

\subsection{D-brane -- anti-D-brane annihilation} 
\begin{figure}
\begin{center}
\includegraphics[width=0.98 \linewidth,keepaspectratio]{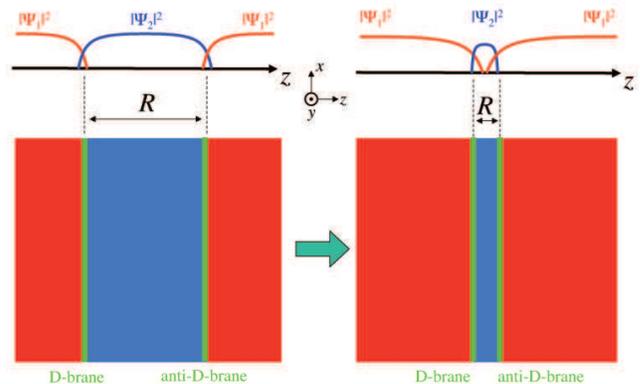}
\end{center}
\caption{Schematic illustration of the D-brane--anti-D-brane collision 
in the BEC system. } 
\label{fig:3} 
\end{figure}

Although brane--antibrane annihilation was demonstrated to show the topological defect creation 
in superfluid $^{3}$He \cite{Bradley}, a physical explanation of the creation mechanism 
of defects still remains unclear. 
We note that the intriguing experiment that mimicked the brane--antibrane annihilation was 
performed by Anderson {\it et al}. \cite{Anderson}.
They created the configuration shown in Fig \ref{fig:3}, 
where the nodal plane of a dark soliton in one component was filled 
with the other component. By selectively removing the filling component 
with a resonant laser beam, they made a planer dark-soliton in a 
single-component BEC. It is known that the planer dark soliton in 3D system 
is dynamically unstable for its transverse deformation 
(known as snake instability) \cite{Anderson}, which results in the decay of the dark soliton 
into vortex rings. 

In our context, this experiment demonstrated the brane--antibrane collision
and subsequent creation of cosmic strings, where the snake instability 
may correspond to "tachyon condensation" in string theory \cite{Sen}. 
The procedure that removes the filling component can decrease
the distance $R$ between two domain walls and cause their collision [see Fig.\ref{fig:3}]. 
The tachyon condensation can leave lower dimensional topological defects 
after the annihilation of D-brane and anti-D-brane. 
In our case of the phase-separated two-component BECs, the annihilation
of the 2-dimensional defects (domain walls) leaves 1-dimensional defects
(quantized vortices),  which we would like to identify
as closed fundamental strings because it can end on a domain wall. 
In string theory, creation of lower dimensional D-branes
after D-brane annihilation was studied very well \cite{Sen},
while creation of closed fundamental strings is in general difficult to deal with.
Contrary to this, it is easy in principle in our case to study closed string creation in
connection with tachyonic fluctuations, which will be one of merits of our system.  

A D-brane soliton similar to ours can be identified
with a D-brane in string theory \cite{Hashimoto:2001rj}
in a tachyon effective field theory (known as the Minahan-Zwiebach model \cite{Minahan:2000ff})
on a non-BPS D-brane. Furthermore, a brane-anti-brane annihilation was studied in
the Minahan-Zwiebach model in \cite{Hashimoto:2002ct}.
There, a string connecting between the two branes
and tachyon fluctuations were studied.
In the case of our D-brane solitons too, one can construct
a string between a pair of brane and anti-brane.
Investigation of tachyonic fluctuations with or without
a string between two branes will be reported elsewhere. 

\subsection{Supersymmetry}
Our system does not have supersymmetry which is a basic ingredient in string theory.
Nevertheless, our D-brane soliton is stable
because it is topological, in contrast to
D-branes in string theory which are stable due to supersymmetry.
As this concern we have two comments.
One is that our system can be made supersymmetric
without changing the bosonic part
by introducing additional fermions to the system. 
For example, the non-relativistic superstring can be realized 
by trapping the fermionic atoms in the core of vortices in a BEC \cite{Snoek}. 
Several studies proposed the possible simulation of (non)relativistic supersymmetry models 
using a mixture of ultracold fermions and diatomic bosons in optical lattices \cite{Yu:2007xb}. 
Then we expect that our D-brane soliton can become a BPS object
preserving a fraction of supersymmetry, as in D-branes in string theory.
In fact at least in the sigma model limit, our solitons reduces to a BPS soliton
of supersymmetric theories \cite{Gauntlett,Shifman,Isozumi,Eto}. 
There is merit in studying further by bringing supersymmetry in our system. 

\acknowledgments
This work was supported by KAKENHI from JSPS
(Grant Nos. 21740267, 199748, 20740141, and 21340104).



\end{document}